\documentclass[%
preprint,
superscriptaddress,
 amsmath,amssymb,
 aps,
prb,
]{revtex4-1}

\usepackage{graphicx}
\usepackage{dcolumn}
\usepackage{bm}

\begin{document}


\title{Observation of spontaneous x-ray magnetic circular dichroism in a chiral antiferromagnet}

\author{Shoya Sakamoto}
\email{These two authors contributed equally to this work}
\affiliation{The Institute for Solid State Physics, The University of Tokyo, Kashiwa, Chiba 277-8581, Japan}

\author{Tomoya Higo}
\email{These two authors contributed equally to this work}
\affiliation{Department of Physics, University of Tokyo, Hongo, Bunkyo, Tokyo 113-0033, Japan}
\affiliation{CREST, Japan Science and Technology Agency (JST), 4-1-8 Honcho Kawaguchi, Saitama 332-0012, Japan}

\author{Masanobu Shiga}
\affiliation{The Institute for Solid State Physics, The University of Tokyo, Kashiwa, Chiba 277-8581, Japan}

\author{Kenta Amemiya}
\affiliation{Institute of Materials Structure Science, KEK, Tsukuba 305-0801, Japan}

\author{Satoru Nakatsuji}
\affiliation{The Institute for Solid State Physics, The University of Tokyo, Kashiwa, Chiba 277-8581, Japan}
\affiliation{Department of Physics, University of Tokyo, Hongo, Bunkyo, Tokyo 113-0033, Japan}
\affiliation{CREST, Japan Science and Technology Agency (JST), 4-1-8 Honcho Kawaguchi, Saitama 332-0012, Japan}
\affiliation{Trans-scale Quantum Science Institute, The University of Tokyo, Bunkyo, Tokyo 113-0033, Japan}
\affiliation{Institute for Quantum Matter and Department of Physics and Astronomy, Johns Hopkins University, Baltimore, Maryland 21218, USA}

\author{Shinji Miwa}
\affiliation{The Institute for Solid State Physics, The University of Tokyo, Kashiwa, Chiba 277-8581, Japan}
\affiliation{CREST, Japan Science and Technology Agency (JST), 4-1-8 Honcho Kawaguchi, Saitama 332-0012, Japan}
\affiliation{Trans-scale Quantum Science Institute, The University of Tokyo, Bunkyo, Tokyo 113-0033, Japan}

\date{\today}

\begin{abstract}
X-ray magnetic circular dichroism (XMCD) signals are usually absent in antiferromagnets. In this letter, we report the observation of spontaneous XMCD spectra originating from the inverse triangular spin structure, or the polarization of the cluster magnetic octupole, in the chiral antiferromagnet Mn$_{3}$Sn thin film. 
The result is consistent with the recent theoretical predictions that the inverse triangular spin structure  can give rise to finite XMCD signals in the absence of net magnetization [J. Phys. Soc. Jpn. 89, 083703 (2020) and Phys. Rev. Lett. 126, 157402 (2021)].

\end{abstract}

\maketitle

\section{Introduction}

\begin{figure*}
\begin{center}
\includegraphics[width=17 cm]{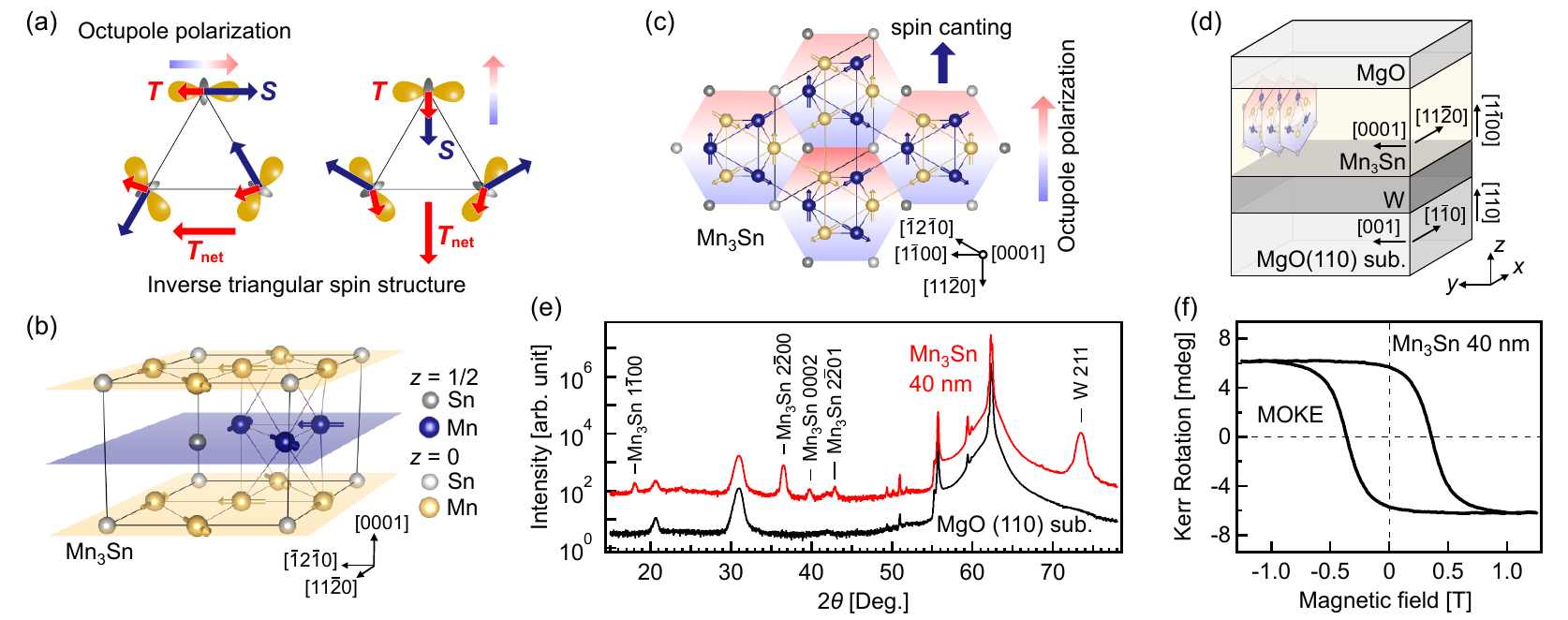}
\caption{(a) Inverse triangular spin ($S$) structure with net magnetic dipole term ($T$).
(b) Crystal structure of $D0_{19}$ Mn$_{3}$Sn. (c) Top view of Mn$_{3}$Sn crystal structure. Mn atoms form two-dimensional Kagome lattices with an inverse triangular spin structure. This spin structure can be viewed as a ferroic order of cluster magnetic octupoles indicated by arrows and hexagons with blue and red gradation. (d) Thin film structure with dominant Mn$_{3}$Sn orientation is indicated.
(e) Out-of-plane x-ray diffraction profiles of the Mn$_{3}$Sn thin film and an MgO(110) substrate. (f) Out-of-plane magnetic hysteresis loop measured using the magneto optical Kerr effect (MOKE).}
\label{Fig1}
\end{center}
\end{figure*}

Antiferromagnets do not possess a net magnetic moment and are hence invisible to common magnetic probes. X-ray magnetic circular dichroism (XMCD) \cite{Stohr:1995ab,Funk:2005aa,van-der-Laan:2014tl}, which typically probes spin and orbital magnetic moments, is usually inapplicable to antiferromagnets in the absence of external magnetic fields.
Recently, Yamasaki {\it et al}. \cite{Yamasaki:2020uy} proposed that the inverse triangular spin (ITS) structure (also referred to as the triangular spin structure with negative chirality) of a noncollinear antiferromagnet can give rise to finite XMCD signals through the non-vanishing magnetic dipole term $T$ even without net magnetization, as schematically shown in Fig. \ref{Fig1}(a). The magnetic dipole term is defined as $T=S - 3(S\cdot\hat{r})\hat{r}$---where $S$ and $\hat{r}$ denote the spin angular momentum and unit electron position operators, respectively---and becomes locally finite when the spin density distribution is anisotropic \cite{Stohr:1995aa}. The spin magnetic moments in the $d_{z^2}$ orbitals are shown as an example in Fig. \ref{Fig1}(a).
Sasabe {\it et al}. \cite{Sasabe:2021ue} extended this idea and showed that the spontaneous XMCD signals can arise even without the local magnetic dipole term as a result of spin- and orbital-direction-dependent x-ray absorption. 

A chiral antiferromagnet Mn$_{3}$Sn with the $D0_{19}$ structure \cite{Nakatsuji:2015aa} is an ideal material for observing the aforementioned XMCD signals originating from the ITS structure because the direction of the ITS structure can be controlled by external magnetic fields, as explained below. Very recently the observation of finite XMCD signals has been reported using single crystalline Mn$_{3}$Sn \cite{Kimata:2021ab}.
Mn$_{3}$Sn consists of the AB stacking of two-dimensional Mn Kagome lattices, depicted by the colored planes in Fig. \ref{Fig1}(b); the inherent geometrical frustration of the Kagome lattice stabilizes the ITS structure, as shown in Fig. \ref{Fig1}(c). 
This spin structure can also be viewed as the ferroic order of the cluster magnetic octupoles \cite{Suzuki:2017vh}, which is represented by hexagons with blue and red colors in Fig. \ref{Fig1}(c). 
Each Mn spin has a local easy axis toward its in-plane nearest-neighbor Sn sites, and only one of the three Mn spins in each Mn triangle is parallel to the local easy axis \cite{Tomiyoshi:1982tn}. As a result, the other two Mn spins show spin canting toward their local easy axis and induce weak ferromagnetism with a net magnetic moment ($\sim 0.002$ $\mu_{\rm B}$/Mn atom) parallel to the octupole polarization \cite{Tomiyoshi:1982tn, Nakatsuji:2015aa}. This weak ferromagnetic moment allows one to control the direction of the octupole polarization by external magnetic fields.

It is worth mentioning that $D0_{19}$ Mn$_{3}$Sn has attracted considerable attention because the octupole polarization bestows Mn$_{3}$Sn with pairs of Weyl nodes in momentum space \cite{Kuroda:2017vl, Yang:2017wi, Chen:2021vu} and induces strong ferromagnetic-like responses such as the anomalous Hall effect \cite{Nakatsuji:2015aa, Zhang:2017tf}, anomalous Nernst effect \cite{Ikhlas:2017aa, Li:2017um}, and magneto-optical Kerr effect \cite{Higo:2018vx,Miwa:2021ua}. 
Octupole polarization switching or manipulation by electric means has also recently been demonstrated \cite{Tsai:2020aa, Tsai:2021vv, Takeuchi:2021vb}.

In this article, we report the observation of XMCD signals originating from the ITS structure, or the polarization of the cluster magnetic octupole, in a Mn$_{3}$Sn thin film with the Kagome planes standing normal to the sample surface. 
The standing Kagome planes allow us to probe the octupole polarization directly.

\section{Methods}
The Mn$_{3}$Sn thin film with the Kagome planes standing normal to the sample surface \cite{You:2019vi, Yoon:2019vw,Takeuchi:2021vb} was grown on a single-crystalline MgO(110) substrate with a W seed layer.
The sample structure, MgO(110) substrate/W (15 nm)/Mn$_{3}$Sn (40 nm)/MgO (3 nm), is schematically shown in Fig. \ref{Fig1}(d).
Mn and Sn atoms were co-deposited using effusion cells.
The sample was annealed at 600 $^{\circ}$C for 30 min after MgO deposition. 
The actual atomic composition was estimated to be Mn$_{3.07}$Sn$_{0.93}$. The detailed sample characterization will be reported elsewhere.

X-ray absorption spectroscopy (XAS) and XMCD measurements were performed on the BL-16A2 beamline at the Photon Factory \cite{Amemiya:2013wl}.
The measurement temperature was room temperature, and magnetic fields ($H$) of up to 5 T were applied. The measurements with 0.1 T and 1 T were performed after applying 2 T along the $z$ direction to saturate the octupole polarization.
The XAS and XMCD spectra were recorded in total electron yield mode. XMCD signals were measured by reversing the helicity of x rays with 10-Hz frequency at each photon energy under a fixed magnetic field \cite{Amemiya:2013wl}.
The measurements were repeated in the opposite magnetic field direction to eliminate artifacts.
The XAS is defined as $(\sigma_{R,+H}+\sigma_{L,+H}) + (\sigma_{R,-H}+\sigma_{L,-H})$, and XMCD is defined as $(\sigma_{R,+H}-\sigma_{L,+H}) - (\sigma_{R,-H}-\sigma_{L,-H})$, where $\sigma_{R/L,\pm H}$ denotes the absorption coefficient measured with right- or left-circularly polarized x-rays under positive or negative magnetic fields.

\section{Results}

Figure \ref{Fig1}(e) shows the out-of-plane $\theta/2\theta$ x-ray diffraction (XRD) patterns of the Mn$_{3}$Sn thin film and MgO(110) substrate. The XRD pattern shows a relatively strong Mn$_{3}$Sn (2$\overline{2}$00) peak and weak (0002) and (2$\overline{2}$01) peaks, indicating that most Kagome planes stand normal to the sample surface. That is, the [$1\overline{1}00$] direction is parallel to the MgO [110] direction, as shown in Fig. \ref{Fig1}(d). 
The XRD $\phi$ scan (not shown) revealed that the Mn$_{3}$Sn [0001] direction was parallel to the MgO [001] direction, and the Mn$_{3}$Sn [$11\overline{2}0$] direction was parallel to the MgO [1$\overline{1}$0] direction, which is consistent with findings of previous studies \cite{Takeuchi:2021vb, Yoon:2021ut}.
Note that the XRD pattern did not show peaks of ferromagnetic WMn$_{2}$Sn impurities, which were reported to appear in W/M$3$Sn heterostructures \cite{Takeuchi:2021vb,Yoon:2021ut}.

The magnetic properties were characterized by magneto-optical Kerr effects (MOKE), and the measured magnetic hysteresis loop at room temperature is shown in Fig. \ref{Fig1}(f). The magnetic fields were applied along the surface normal direction. 
The MOKE showed hysteresis with a coercive field of 0.35 T, which is larger than the bulk value of $\sim$0.05 T \cite{Nakatsuji:2015aa} but smaller than the value for randomly-oriented polycrystalline thin films, 0.6 T \cite{Higo:2018ud}.
The sizable Kerr rotation most likely originates from the octupole polarization in Mn$_{3}$Sn \cite{Higo:2018vx, Miwa:2021ua} because the sign of the Kerr rotation angle was opposite to that of typical ferromagnets such as Fe and because the ferromagnetic moment induced by spin canting ($< 10$ m$\mu_{\rm B}$/Mn atom, as described below) is almost negligible.

\begin{figure}
\begin{center}
\includegraphics[width=8.1cm]{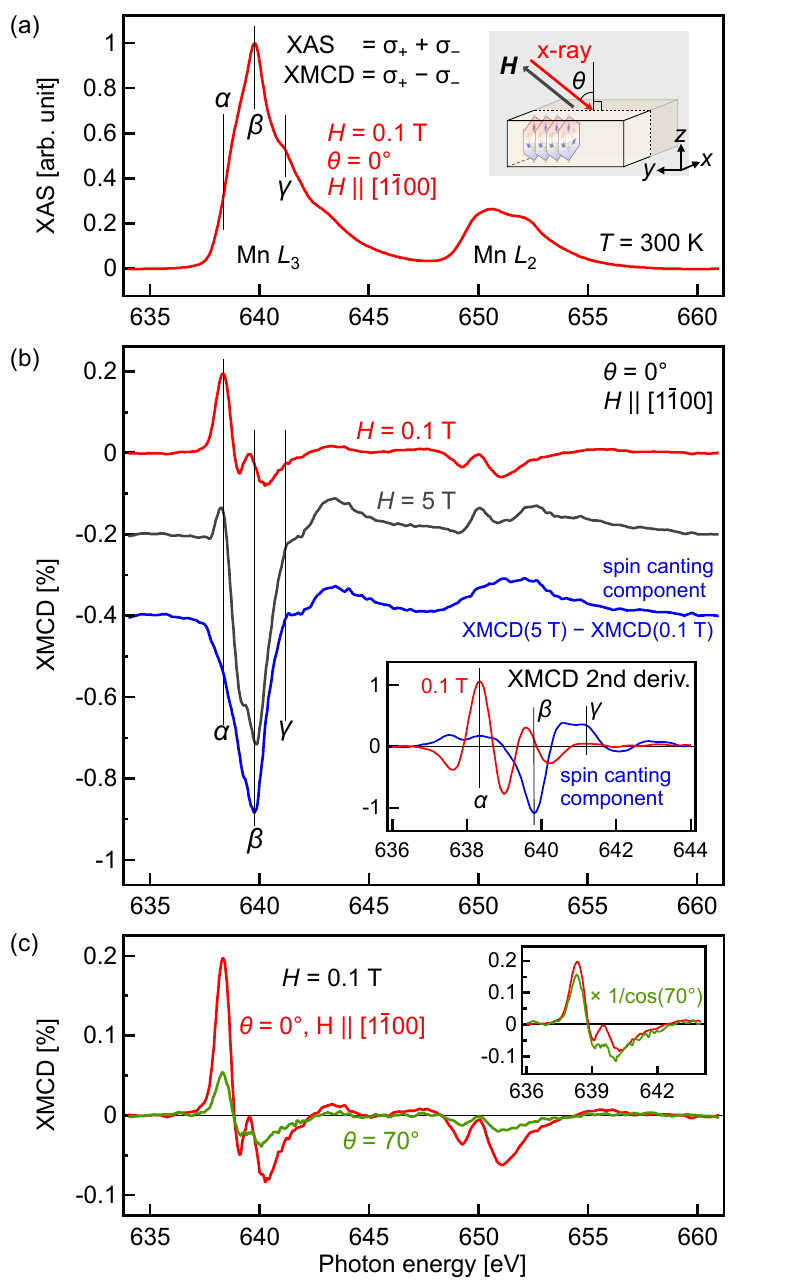}
\caption{(a) Typical XAS spectrum of the Mn$_{3}$Sn thin film. The inset shows the measurement geometry. (b) XMCD spectra taken with $H$ = 0.1 T and $H$ = 5 T under the normal-incidence (NI) geometry with $\theta=0^{\circ}$. The spin canting component (blue curve) is extracted by subtracting the 0.1 T spectrum from the 5 T spectrum. The second derivative spectra are shown in the inset. (c) XMCD spectra taken under the NI geometry and the grazing-incidence geometry ($\theta=70^{\circ}$). The inset shows the magnified view around the Mn $L_{3}$ edge.}
\label{Fig2}
\end{center}
\end{figure}

Figure \ref{Fig2}(a) shows a typical XAS spectrum of the Mn$_{3}$Sn thin film. The background, consisting of a double step function and a linear function, is subtracted. The XAS spectral line shape did not depend on the magnetic field strength or direction. The inset of Fig. \ref{Fig2}(a) shows the measurement geometry, where the incident x ray is in the $yz$ plane, and the angle between the incident x ray and the sample normal direction is denoted by $\theta$. 

The XMCD spectra obtained with $H$ = 0.1 T and 5 T in the normal incidence (NI) geometry ($\theta = 0^{\circ}$) are shown in Fig. \ref{Fig2}(b). The spectra before performing magnetic-field average are shown in the Supplementary material \cite{SupplMn3SnXMCD}.
The spectral line shapes significantly differ from each other: the spectrum obtained with $H$ = 0.1 T shows a strong positive pre-edge peak, denoted by $\alpha$ in Fig. \ref{Fig2}(b), while the spectrum obtained with $H$ = 5 T shows an intense negative peak at the XAS peak position, denoted by $\beta$ in Fig. \ref{Fig2}(b).
The XMCD spectra are expected to consist of the spin-canting component, which develops with increasing magnetic fields, and the field-independent octupole component.
The peak $\beta$ most likely originates from the spin-canting component, as the peak develops with increasing magnetic fields.
If the spectrum taken with $H$ = 0.1 T is subtracted from the spectrum taken with $H$ = 5 T, a broad featureless XMCD spectrum without peak $\alpha$ is obtained, as shown by the blue curve in Fig. \ref{Fig2}(b). 
This broad spectrum resembles the XMCD spectra of metallic systems, suggesting that it reflects the spin component. Hereafter, we refer to this difference spectrum as the spin-canting XMCD spectrum.
It follows that the spectrum taken with $H$ = 0.1 T, which shows multiple sharp peaks, predominantly reflects the octupole component.
Note that a previous study on a Mn$_{3}$Sn ($40\overline{4}3$) thin film observed field-induced XMCD signals, the magnitude of which is comparable to that obtained in the present study, but the pre-edge peak $\alpha$, a signature of the octupole polarization, was not observed \cite{Taylor:2020uf}.

To examine the spectral line shapes in more detail, the second derivatives of the XMCD spectra are shown in the inset of Fig. \ref{Fig2}(b). 
The second derivative of the spin-canting XMCD spectrum does not exhibit a peak at position $\alpha$, indicating that the XMCD signals originating from the octupole polarization do not depend on the magnetic field strength.
In contrast, the XMCD spectrum obtained with $H$ = 0.1 T is expected to contain a finite spin-canting component in addition to the octupole component. To estimate how much the spin-canting component contributes to the spectrum taken with $H$ = 0.1 T, we subtracted a fraction of the spin-canting XMCD spectrum from the $H$ = 0.1 T spectrum and minimized the kink feature $\gamma$ around 642 eV and the main feature $\beta$. 
From this analysis, it was determined that approximately $13 \pm 13$ \% of the spin canting component contributes to the XMCD spectrum taken with $H$ = 0.1 T.
The XMCD spectrum that only reflects the octupole polarization can thus be obtained as ${\rm XMCD(0.1\ T)} - 0.13\times {(\rm spin}$-canting XMCD), which we refer to as the octupole XMCD.

The XMCD spectra captured in the NI geometry ($\theta=0^{\circ}$) and the grazing-incidence (GI) geometry ($\theta = 70^{\circ}$) with a magnetic field of 0.1 T are shown in Fig. \ref{Fig2}(c). Here, the probing depth in the GI geometry is almost the same as that in the NI geometry because a typical x-ray penetration depth of $\sim$ 100 nm is much longer than a typical electron escaping depth of $\sim$ 5 nm.
The XMCD signals captured in the GI geometry are smaller than those captured in the NI geometry, and they seem to scale with $\cos \theta$, as can be seen in the inset of Fig. \ref{Fig2}(c). This scaling behavior suggests that the XMCD signals at 0.1 T originate from the octupole polarization, which always lies in the Kagome plane, projected onto the x-ray incidence axis. The small difference in the spectral line shape may have originated from the spin-canting component.
Note that the XMCD spectrum (not shown) taken with the magnetic field along the [$\overline{1}2\overline{1}0$] direction (30-degree away from the sample normal in the $xz$ plane) looked almost the same as that taken in the NI geometry as expected \cite{Yamasaki:2020uy, Sasabe:2021ue}.

Here, we estimate the spin and orbital magnetic moments using the XMCD sum rules \cite{Thole:1992aa, Carra:1993aa, Chen:1995aa}:  
\begin{eqnarray}
&\displaystyle m_{\rm orb}=-\frac{4}{3}\frac{S^{\rm XMCD}_{L_{3}} + S^{\rm XMCD}_{L_{2}}}{S^{\rm XAS}}n_{h},\\
&\displaystyle m_{\rm spin}^{\rm eff} = m_{\rm spin}+7m_{\rm T}=-2\frac{S^{\rm XMCD}_{L_{3}} - 2 S^{\rm XMCD}_{L_{2}}}{S^{\rm XAS}}n_{h},
\end{eqnarray}
where $m_{\rm orb}$ and $m_{\rm spin}$ are the orbital and spin magnetic moments in units of $\mu_{\rm B}$/atom, respectively, and $\mu_{\rm B}$ is the Bohr magneton; $m_{\rm T}$ is the expectation value of the magnetic dipole term $m_{\rm T} = -\langle T \rangle \mu_{\rm B}/\hbar$, with $\hbar$ being the reduced Planck constant \cite{Stohr:1999aa}; $S^{\rm XMCD}_{L_{2,3}}$ represents the XMCD integral over the $L_{2}$ or $L_{3}$ edges, and $S^{\rm XAS}$ represents the XAS integral over the $L_{2,3}$ edges; $n_{h}$ is the number of 3$d$ holes, which was assumed to be five in the present study.

\begin{figure}
\begin{center}
\includegraphics[width=8.3 cm]{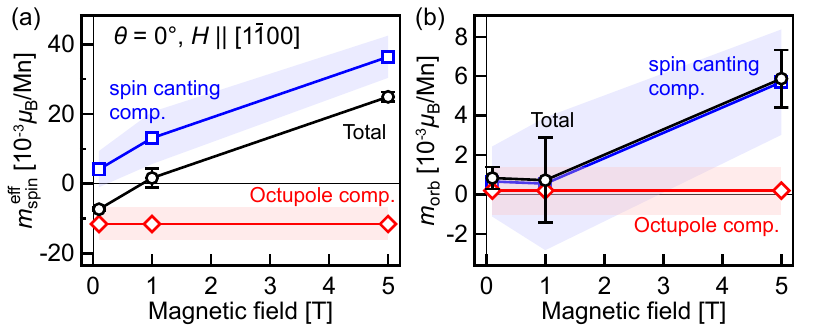}
\caption{Magnetic field dependence of (a) the effective spin magnetic moment and (b) the orbital magnetic moment deduced using the XMCD sum rules in the NI geometry (black circles). The black error bars represent uncertainty in the spectral background subtraction procedure. The data are decomposed into the spin canting component (blue squares) and octupole component (red diamonds). The shaded areas represent the uncertainty in the spectral decomposition procedure.}
\label{Fig3}
\end{center}
\end{figure}

The deduced effective spin magnetic moments $m_{\rm spin}^{\rm eff}$ and orbital magnetic moments $m_{\rm orb}$ are plotted with black circles in Figs. \ref{Fig3}(a) and \ref{Fig3}(b), respectively.
The effective spin magnetic moment increases as the magnetic field increases, and changes the sign at around 1 T. 

The spectral decomposition procedure mentioned above can separate the deduced magnetic moments into the spin canting component and the octupole component, as shown by the blue squares and red diamonds, respectively, in Figs. \ref{Fig3}(a) and \ref{Fig3}(b).
Here, we assume that the octupole component does not depend on the magnetic field, which is evidenced by the fact that the spin-canting XMCD spectrum does not show any octupole XMCD features.
The effective spin magnetic moment, or the magnetic dipole term, of the octupole component is found to be antiparallel to that of the spin-canting component.
These negative field-independent effective spin magnetic moments indicate that the octupole XMCD indeed originates from the inverse triangular spin structure of Mn$_{3}$Sn because if the octupole XMCD originated from ferromagnetic or superparamagnetic impurities, the effective spin magnetic moments would have been positive and magnetic field dependent.
The orbital magnetic moment is almost absent for the octupole component but is finite for the spin-canting component. The absence of the orbital magnetic moment is consistent with the fact that the magnetic octupole polarization does not possess a net magnetic moment.

The decomposition analysis showed that the spin canting component was no more than 0.04 $\mu_{\rm B}$/Mn—even at 5 T—and $0.004 \pm 0.005$ $\mu_{\rm B}$/Mn in the remanence state. This almost negligible remanent spin magnetic moment is consistent with the bulk value of 0.002 $\mu_{\rm B}$/Mn \cite{Nakatsuji:2015aa}. 
Despite this consistency, there still is a possibility that the tiny field-dependent spin magnetic moments were induced by minor paramagnetic or ferromagnetic second phases or the spin canting of Mn$_{3}$Sn grains with different orientations.
Note that the magnetic moments are expected to be slightly underestimated because of the existence of a small number of Mn$_{3}$Sn grains with different orientations and the spectral overlap between the Mn $L_{2,3}$ edges \cite{Teramura:1996aa}.

\begin{figure}
\begin{center}
\includegraphics[width=7.16 cm]{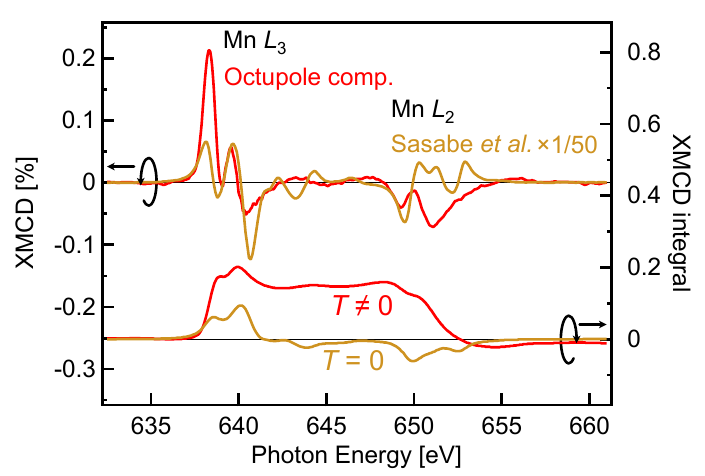}
\caption{Comparison between the octupole XMCD spectrum (red) and the theoretical spectrum (orange) \cite{Sasabe:2021ue}. The integral XMCD spectra are also shown at the bottom.}
\label{Fig4}
\end{center}
\end{figure}

\section{Discussion}

Figure \ref{Fig4} compares the octupole XMCD spectrum obtained in this study with a previously reported theoretical spectrum \cite{Sasabe:2021ue}.
The overall line shapes roughly agree with each other, especially for the Mn $L_{3}$ edge.
However, the XMCD signals were more than an order of magnitude weaker than those of the calculations. This discrepancy may be due to the fact that the theoretical calculations assumed an isolated ionic Mn$^{2+}$ high-spin state with $\langle S \rangle/\hbar = 5/2$, $\langle L \rangle/\hbar \sim 0$, $\langle T \rangle/\hbar \sim 0$, which is quite different from the local electronic structure of the Mn atoms in Mn$_{3}$Sn with a metallic nature and a local magnetic moment of $\sim3$ $\mu_{\rm B}$ \cite{Brown:1990tm}.

In theoretical calculations, it was argued that the sizable XMCD signals arise even without the net spin magnetic moment or net magnetic dipole term. This can be seen in the XMCD integral becoming zero above both the Mn $L_{3}$ and $L_{2}$ edges ($S^{\rm XMCD}_{L_{3}}=S^{\rm XMCD}_{L_{2}} = 0$).  
The finite oscillating XMCD spectrum can be considered to originate from the magnetic dipole term of each Mn 3$d$ orbital, which is split in energy by a crystal field and has a different x-ray absorption process. For example, when the occupied $d_{z^{2}}$ and $d_{x^{2}-y^{2}}$ orbitals with the opposite magnetic dipole terms are non-degenerate, they would exhibit XMCD spectra that are different in line shape and hence a finite superimposed XMCD spectrum regardless of whether or not net magnetic dipole term exists.

Experimentally, the XMCD integral over the Mn $L_{3}$ edge remains finite and approaches zero above the Mn $L_{2}$ edge ($S^{\rm XMCD}_{L_{3}}\neq0$ and $S^{\rm XMCD}_{L_{3}} + S^{\rm XMCD}_{L_{2}}=0$). 
As shown in Figs. \ref{Fig3}(a) and \ref{Fig3}(b), this integral results in a finite effective spin magnetic moment but a negligible orbital magnetic moment.
We attribute the origin of the finite effective spin magnetic moment to the non-vanishing magnetic dipole term $T$ in the cluster magnetic octupole.
Note that the finite net magnetic dipole term does not seem necessary to observe XMCD signals as described above.

The present findings demonstrate that non-collinear antiferromagnetic spin structures can be probed by XMCD. This enables the resolution of antiferromagnetic domains and antiferromagnetic domain walls using the scanning XMCD technique \cite{Fischer:2015tw, Kotani:2018tj}. 
The local anisotropic spin distribution may also be elucidated by analyzing the spectral line shape in greater detail \cite{Shibata:2018aa,Sakamoto:2021wq}.

\section{Summary}
In this letter, we report the observation of XMCD signals arising from the non-collinear antiferromagnetic spin structure of a Mn$_{3}$Sn thin film with the Kagome planes standing normal to the sample surface. We reveal that the finite net magnetic dipole term in the inverse triangular spin structure, or the cluster magnetic octupole, induces the XMCD signals.

\newpage
\section*{Supplementary Information}
\subsection{XMCD spectra recorded with positive and negative magnetic fields}

\begin{figure*}[h]
\begin{center}
\includegraphics[width=9 cm]{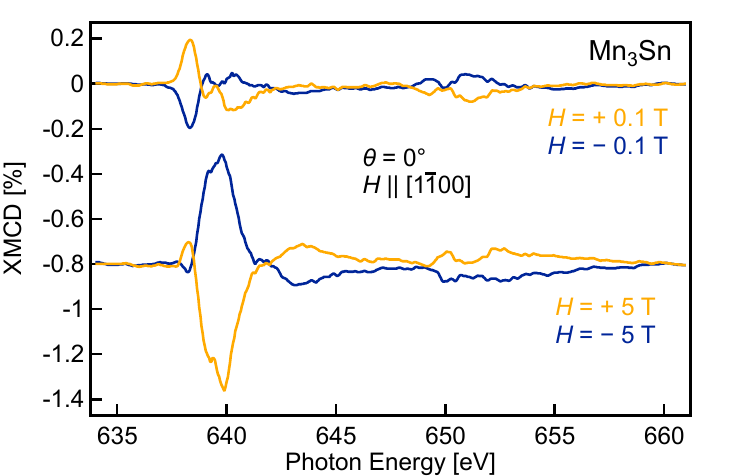}
\caption{XMCD spectra of the Mn$_{3}$Sn thin film recorded with $\pm 0.1$ T and  $\pm 5$ T.}
\label{FigS1}
\end{center}
\end{figure*}

Figure \ref{FigS1} shows the x-ray magnetic circular dichroism (XMCD) spectra captured with positive and negative magnetic fields. The XMCD spectra reversed their signs with the magnetic-field directions, which rules out the possibility that the XMCD signals originated from non-magnetic artifacts. For example, when photon energies of right- and left-circularly polarized x-rays are slightly different, the photon-energy derivative of absorption spectra can be obtained. Such artifact can be removed by averaging XMCD spectra taken with positive and negative magnetic fields, as shown in the main text.

\begin{acknowledgments}
We thank K. Kondou of RIKEN for discussion. This work was performed under the approval of the Photon Factory Program Advisory Committee (Proposal No. 2019S2-003).
This work was supported by JSPS KAKENHI (Nos. JP18H03880, JP19H00650, JP20K15158), JST CREST (JPMJCR18T3), JST-Mirai Program (JPMJMI20A1), and Spintronics Research Network of Japan (Spin-RNJ). Institute for Quantum Matter, an Energy Frontier Research Center was funded by DOE, Office of Science, Basic Energy Sciences under Award No. DE-SC0019331.
\end{acknowledgments}

\bibliography{BibTex_all}
\end{document}